\documentclass[amsmat,amssymb,amsfonts,aps,prb,twocolumn,showpacs,superscriptaddress]{revtex4}

\usepackage{graphicx}
\usepackage{dcolumn}
\usepackage{bm}
\usepackage{color}

\begin{document}

\title{Anisotropic magnetoresistance in nanocontacts}

\author{D. Jacob}
\affiliation{Department of Physics and Astronomy, Rutgers University, Piscataway, NJ 08904, USA}
\affiliation{Departamento de F{\'\i }sica Aplicada, Universidad de 
Alicante, San Vicente del Raspeig 03690, Alicante, Spain}
\author{J. Fern\'andez-Rossier}
\affiliation{Departamento de F{\'\i }sica Aplicada, Universidad de 
Alicante, San Vicente del Raspeig 03690, Alicante, Spain}
\author{J. J. Palacios }
\affiliation{Departamento de F{\'\i }sica Aplicada, Universidad de 
Alicante, San Vicente del Raspeig 03690, Alicante, Spain}
\date{\today}

\begin{abstract}
We present {\it ab initio} calculations of the evolution of
anisotropic magnetoresistance (AMR) in Ni nanocontacts
from the ballistic to the tunnel regime. We find an 
extraordinary enhancement of AMR, compared to bulk, 
in two scenarios. In systems without localized states, 
like chemically pure break junctions, 
large AMR only occurs if the orbital polarization of the 
current is large, regardless of the 
anisotropy of the density of states. In systems that display 
localized states close to the Fermi energy, like a single 
electron transistor with ferromagnetic electrodes, large AMR 
is related to the variation of the Fermi energy as a function 
of the magnetization direction.
\end{abstract}

\maketitle
	

\section{Introduction}

The free energy of mono-domain ferromagnetic particles depends on the 
relative orientation of the magnetization with respect to the crystal 
lattice. This magnetic anisotropy results from the combination of 
Coulomb repulsion favoring spin polarization,  spin-orbit coupling
(SOC), and the crystal field breaking the orbital rotation invariance. 
As a result, the orbital moment of magnetic atoms and their magnetic 
anisotropy energy (MAE) depend strongly on their atomic coordination 
\cite{Gambardella03,Canali07}. 

The transport counterpart of MAE is 
anisotropic magneto-resistance (AMR), i.e. the dependence of the resistance 
on the angle $\theta$ between the magnetization and the current flow. 
Whereas AMR in bulk was known back in the XIX century and is a rather small 
effect, the recent observation of AMR in a variety of low dimensional systems 
\cite{Gould04,Ruster05,Giddings05,Saito07,Moser07,Grigorenko06,Natelson-BAMR,Ralph-BAMR,Viret-BAMR,Nature07}, largely exceeding 
bulk values, has opened a new research venue in the field of spin-polarized 
quantum transport. Very large AMR  has been reported in  planar tunnel
junctions  (TAMR) with a variety of electrode and barrier materials
\cite{Gould04,Ruster05,Giddings05,Saito07,Moser07,Grigorenko06}. Enhanced  AMR has also been observed in atomic sized contacts,
both in the tunnel regime (TAMR) and in the contact (or ballistic\cite{ballistic}) 
regime (BAMR)\cite{Velev05}, for Py \cite{Ralph-BAMR}, Fe  \cite{Viret-BAMR}, Ni
\cite{Natelson-BAMR}, and Co \cite{Nature07}. Additionally,  GaMnAs islands in
the Coulomb Blockade regime show electrically tunable AMR (CB-AMR) 
\cite{Wunderlich}. Thus, a wide range of nanostructures made from different
materials  display enhanced AMR.

Here we focus on AMR in atomic-size conductors for several reasons. On the one hand, 
conductance of atomic-sized contacts probes the electronic structure of the apex atoms. 
These have coordination different from bulk and thus present different orbital and spin 
magnetic moment \cite{Pt:2005}, and enhanced magnetic anisotropy 
\cite{Gambardella03,Canali07,Bruegel06} which might be probed by BAMR. On the other hand, 
nanocontacts allow to study AMR going from the contact (BAMR)  to the tunnel (TAMR) regime 
in the same system, as shown in the case of both Ni and Py \cite{Ralph-BMR,Ralph-BAMR}. Ni 
nanocontacts have also been used as electrodes to explore the Coulomb Blockade and the 
Kondo regimes \cite{Ralph-Science}.

The crux of the matter is to identify the necessary and sufficient conditions 
to expect large values of AMR in quantum transport. Here we consider two
different  transport regimes, coherent and sequential. 
In the coherent regime we use the Landauer formalism that, at zero temperature, 
relates the zero-bias conductance $G$ to the quantum mechanical transmission of the electrons at the
Fermi  energy, $G=\frac{e^2}{h}T(\epsilon_F)$. This approach accounts for AMR
both in the tunneling regime (TAMR)\cite{TAMR-theory} and in the contact or
ballistic regime (BAMR)\cite{Velev05}  in the absence of sharp resonances near
the Fermi energy. In the scattering-free case of  perfect 1D chains,
$T(\epsilon_F)$ is simply given by the number of bands at the Fermi  energy
${\cal N}(\epsilon_F)$. Because of the SOC, ${\cal N}(\epsilon_F)$ for
ferromagnetic  1D transition metal chains
\cite{Velev05,Viret-BAMR,Bruegel06,Nature07} depends on the angle  $\theta$
between the chain axis and the magnetization, and this leads naturally to
stepwise  $G(\theta)$ curves.

However, the idealized scattering-free picture fails to account for the 
experimental results of conductance in metallic nanocontacts, for which 
scattering channels are not perfectly conducting \cite{NMNC}. According 
to the scattering-free theory, the conductance of atomic contacts of Ni, 
in units of $e^2/h$, would be 6 or 7 for Ni \cite{Velev05}, in quantitative 
{\em disagreement} with the measured \cite{Untiedt:prb:04} conductance 
of Ni nanocontacts around $3 e^2/h$. The same applies to Fe, Co and Pt. Scattering  
definitely affects $d$-bands, which suffer the so called {\it orbital blocking}
\cite{Jacob05}. 

Here we present calculations of BAMR {\em plus} scattering. This 
approach also permits to calculate the crossover from BAMR to TAMR. We find that
in the  coherent regime large AMR is related to the orbital polarization of the
current. TAMR  has been linked to the anisotropy of the density of states at
$\epsilon_F$, which turns  out to be large in Ni chains. Unexpectedly, this does
not lead to a large value of TAMR,  the reason being that the current is not
orbitally polarized in this limit. 

In the sequential regime, valid to describe systems that feature transport
through resonant  levels of width $\Gamma$ smaller than the temperature
$kT$\cite{Been}, we find enhanced AMR,  regardless of the orbital polarization
of the current, if the chemical potential $\epsilon_F$  of the ferromagnetic
electrode crosses a resonance as it varies due to change of the magnetization 
angle. This situation occurs in a single electron transistor with ferromagnetic
electrodes \cite{Wunderlich,CBAMR-theory}. Resonances might also occur in the
tip atoms of Ni nanocontacts in the tunneling regime\cite{Burton07}.

This paper is organized as follows: In Sec. \ref{sec:model} we introduce the 
model system and the theoretical method for the calculation of AMR in magnetic
nanocontacts. In Sec. \ref{sec:contact} we calculate the AMR in the contact or 
ballistic regime (BAMR). In Sec. \ref{sec:tunnel}
we treat AMR in the {\it coherent} tunneling regime (TAMR), while in Sec. 
\ref{sec:sequential} we treat AMR in the {\it sequential} tunneling regime.
Finally, in Sec. \ref{sec:summary} we conclude the paper summarizing the main 
results. 

\section{Model and Methodology} 
\label{sec:model}

\begin{figure}[t]
  \includegraphics[width=\linewidth]{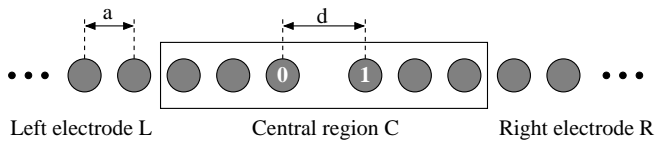}
  \caption{
   1D Model of Ni nanocontact: Elastic electron scattering in the contact
    region of a real  nanocontact is mimicked by a gap $d>a$ between two
    semi-infinite 1D Ni chains with lattice  spacing $a$.
  }
  \label{fig:model}
\end{figure}

As a model system for ferromagnetic nanocontacts we consider two semi-infinite
Ni 1D chains  with lattice parameter $a$, separated by a gap $d>a$, as shown in
Fig. \ref{fig:model}. This  model shares most of the relevant features with
realistic nanocontact models, like e.g. the  low coordination of the tip atoms
of the two electrodes and elastic electron scattering due  to the gap. On the
other hand the one-dimensionality and the resulting rotational invariance of our
model simplify considerably the calculations of the SOC and the interpretation
of the results. Such one-dimensional models have been employed before to study
fundamental properties of atomic-size nanocontacts
\cite{Delin,Smogunov,DalCorso}.
 
We calculate the electronic structure of the system using a combination of density functional
theory (DFT) in the local spin density approximation (LSDA) and a Green's function technique 
to account for the fact that, when $d\neq a$, the system is not translationally invariant. We 
split the system into 3 regions, left (L) and right (R) electrodes, described as semi-infinite 
Ni chains, and the central region (C) containing the 3 innermost atoms of each electrode, as 
shown in Fig. \ref{fig:model}. 

The electronic structure of both the electrodes and the central region is
described using  effective one-body Hamiltonians obtained from {\it ab initio}
calculations, performed with  CRYSTAL03\cite{CRYSTAL03} on the LSDA level, and
using a localized atomic orbital minimal basis set. CRYSTAL03, which does not
include SOC, yields spin-polarized solutions along an arbitrary  axis with
majority and minority electrons. The SOC term $\hat{H}_{\rm
SO}=\lambda\vec{L}\cdot\vec{S}$ is added to the converged self-consistent LSDA
Hamiltonian $\hat{H}_{\rm LSDA}$:
\begin{equation}
  \label{eq:hamiltonian}
  \hat{H} = \hat{H}_{\rm LSDA} + \hat{H}_{\rm SO}.
\end{equation}
This post self-consistent approach \cite{Jaime,Velev05} is justified in the case
of Ni, for  which the SOC is much smaller than the exchange splittings and the
bandwidths. We take  $\lambda=70 meV$ for the Ni $3d$-orbitals.

The Green's function of the central region is obtained by means of the
so-called {\it partitioning} technique:
\begin{equation}
  \label{eq:GC}
  \hat{G}_C(E) = (E-\hat{H}_C-\hat\Sigma_L(E)-\hat\Sigma_R(E))^{-1},
\end{equation}
where $\hat{H}_C$ is the total Hamiltonian (including SOC) of the C region, and
$\hat\Sigma_L$ and  $\hat\Sigma_R$ are self-energies that take into account the
coupling of the central region to the  two electrodes
\cite{Datta,Jacob:thesis}. 

In the coherent regime, i.e. at low temperatures and small bias 
voltages when inelastic scattering events can be neglected, we use
the Landauer formalism to calculate the conductance of the system
which is obtained from the transmission function.
The transmission function in turn can be calculated by means of the 
Caroli expression from the Green's function of the central 
region, eq. (\ref{eq:GC}), and the so-called coupling matrices,
$\hat\Gamma_L(E):=i(\hat\Sigma_L-\hat\Sigma_L^\dagger)$ and 
$\hat\Gamma_R(E):=i(\hat\Sigma_R-\hat\Sigma_R^\dagger)$, of 
the electrodes \cite{Caroli}:
\begin{equation}
  \label{eq:transm}
  T(E) = {\rm Tr\,}[ \hat{G}_C(E) \, \hat\Gamma_L(E) \, \hat{G}_C^\dagger(E)
   \, \hat\Gamma_R(E) ].
\end{equation}
The zero-bias conductance is then given by $G=\frac{e^2}{h}T(\epsilon_F)$. The
orbital projected density of states $\rho_\alpha$ and the density of states of
the central region $\rho$ can be calculated from the Green's function:
$\rho_\alpha(E)=-\frac{1}{\pi}{\rm Im}[G_{\alpha\alpha}(E)]$
and $\rho(E)=-\frac{1}{\pi}{\rm Im \, Tr}[\hat{G}_C(E)]$.

\section{Contact regime}
\label{sec:contact}
\begin{figure}
[hbt]
\includegraphics[width=\linewidth]{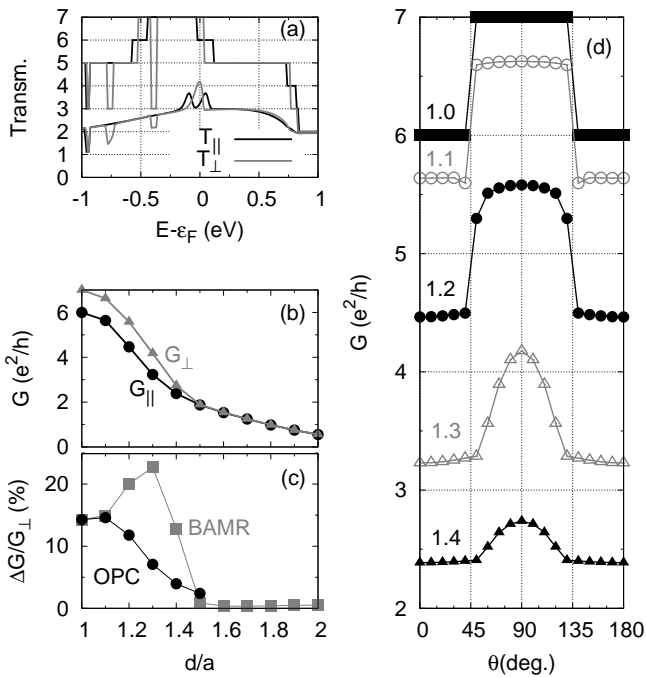}
\caption{
  (a) Transmission for an ideal infinite Ni chain ($d=a$, upper curves) 
  and for two semi-infinite Ni chains seperated by $d=1.3a$ (lower curves) for 
  magnetization parallel (black) and perpendicular (grey) to the chain axis. 
  (b) Zero-bias conductance $G$ as a function of $d$ for magnetization parallel 
  (black) and perpendicular (grey) to the chain axis. (c) BAMR (grey boxes) and OPC 
  (black circles) as a function of $d$. (d) $G$ as a function of 
  $\theta$ for different values of $d/a$: 1.0 (full black boxes), 1.1 (grey
  circles), 1.2 (black full circles), 1.3 (grey triangles) and 1.4 (black 
  full triangles).
}
\end{figure}

\subsection{The ideal chain}
Magnetic anisotropy comes from the combined action of both, the crystal field that 
breaks the orbital rotational invariance, and the atomic SOC term, that couples the
spin polarization to the orbital degrees of freedom. 
The electronic structure of the ideal one dimensional Ni chain ($d=a$) presents
a number  of common features with 3d and 4d transition
metals\cite{Bruegel06,Wierzbowska:prb:05}  and permits to understand  transport
results for $d\neq a$. The bands close to the Fermi  energy are formed by  $s$
and $d$ orbitals. In the absence of SOC, rotational invariance  around the chain
axis permits to classify the $d$ orbitals according to the projection of  their
angular momentum along the chain direction, $m_z$. On top of that,  a weak
crystal  field splits the otherwise degenerate $d$ levels into two doublets
$E_1$ (linear combination of states with $m_z=\pm 1$) and $E_2$  ($m_z=\pm 2$)
and a singlet  $A_1$  ($m_z=0$), which  is hybridized with the $s$ orbital. The
orbital degeneracy of the doublets is kept by the  bands of the chain, as long
as SOC is not present. The bandwidth of the $E_2$ is significantly  smaller than
that of $E_1$ due to the smaller overlap of the $E_2$ orbitals. As a result, 
the $E_2$ bands yield a higher density of states. 
 
The combined action of SOC and magnetism alters this situation
\cite{Velev05,Bruegel06}.   When the magnetization is pointing perpendicular to
the chain axis $(\theta=90^\circ)$,  SOC acts as an effective  magnetic field
acting over $L_x$ that has to compete with the  $L_z^2$-like terms arising from
the crystal field, which happens to be dominant. As a  result, the bands for
$\theta=90^\circ$ look very similar to those without SOC, except  in the points
where bands with $m_z \sigma $ and $m_z\pm1,\sigma\mp 1$ intersect, which are  far
away from $\epsilon_F$ in the case of Ni. Therefore, for $\theta=90^\circ$ the 
effect of SOC on transport is negligible. In contrast, when magnetization is
pointing  along the chain axis $(\theta=0^\circ)$, SOC shifts the bands by an
amount  $\lambda m_z \sigma$,  where $m_z$ and $\sigma$ are the projection of
the spin and  orbital momentum along the chain. As a result, the $E_2$ and $E_1$
orbital doublets are  split so that one of the 2 minority  $E_2$ bands is
shifted below the Fermi energy,  compared to the $\theta=90^\circ$ case. This 
can be seen in the stepwise curves in Fig.  2a, that  correspond to 
$T_{\|}(E)\equiv T(E,\theta=0^\circ)$ and  $T_\bot(E)\equiv
T(E,\theta=90^\circ)$ for the ideal chain. At the Fermi energy,  $T_\bot(E)\neq
T_\|(E)$. This change is responsible for BAMR\cite{Velev05}, defined as
BAMR$\equiv \frac{\Delta G}{G_\bot}\times 100$ where $\Delta G\equiv
G_\bot-G_\|$.

The interplay between SOC and magnetization results in a non-zero orbital
moment  {\em density} along the magnetization direction. 
The largest orbital moment occurs when  $\theta=0$, i.e. when the magnetization is
along the chain \cite{Bruegel06}.  The 
{\em orbital} {\em polarization current} (OPC) defined as 
\begin{equation}
  {\rm OPC} \equiv \frac{\sum_{m} T_m-T_{-m}}{T(E)}
\end{equation}
where $T_m$ is the transmission of the $d$-orbitals with $m=\pm 2$ or $m=\pm 1$
along the  chain direction, vanishes when $\theta=90^\circ$ but  is {\em
non-zero} when  $\theta=0$. Interestingly, there is a perfect one-to-one
correspondence between the OPC and the BAMR in the case of the ideal chain
without scattering. It is also apparent that  the existence of an orbital
magnetic moment is not a sufficient condition for having a  non-zero OPC, very
much like spin-polarization does not necessarily imply a spin-polarized  current
\cite{Jacob05}.

\subsection{The effect of weak scattering }

Now we see how elastic scattering, controlled with the chain separation $d$,
affects BAMR. The  stretched bond mimics the contact region. This perturbation
preserves the axial symmetry  of the ideal chain but introduces scattering. As a
consequence $T(E)$, shown in Fig. 2a,   is not quantized anymore, as expected
\cite{Jacob05,Untiedt:prb:04} and yet the BAMR  (Fig. 2c) is close to that of
the ideal case for values of $d/a\le1.4$. Relatedly, the  $G(\theta)$ curve is
not stepwise (as is the case of the ideal chain) anymore when scattering  is
included (Fig. 2d). On the other hand the $G(\theta)$ curve is also different
from bulk  behavior where $G(\theta)\propto\cos^2\theta$. The quantized step in
the ideal case ($d=a$)  that corresponds to the critical angle at which the $E_2$
band is pushed below the Fermi energy  \cite{Velev05}, becomes progressively
smoother as the gap between the chains increases. Our  $G(\theta)$ curves
including scattering agree with those of the experiments \cite{Viret-BAMR}. 
This is one of the important results of the model. 

As $d$ increases, the scattering increases and $G$ goes down but interestingly, the BAMR 
signal first \emph{increases} slightly for $d/a\le1.3$ before finally going down with 
increasing scattering. The initial increase in BAMR is related to the initially stronger 
decrease of the contribution of the $A_1$-channel to the conductance. The $A_1$-channel is 
not affected by the SOC in contrast to the $E_2$ channels that are mainly responsible 
for the BAMR signal. The decrease of the BAMR signal for larger values of $d$ is expected 
within the framework of our model, since the relative contribution to the conductance of 
the $d$-channels compared to the $s$-channel decreases as the gap opens. The reason for 
this is the shorter spread of the $d$-orbitals compared to the $s$-orbitals \cite{Jacob05}. 
Relatedly, the OPC (Fig. 2c) also decreases as $d$ increases. Removing the contribution of 
the $s$-channel to the conductance would thus enhance BAMR. This could be accomplished e.g. 
by oxidation of the contact \cite{Jacob06}.

\section{Coherent tunneling regime}
\label{sec:tunnel}

In this section we study  the anisotropic magnetoresistance in the regime of weakly coupled 
semi-infinite chains. In Figs. 3a and 3c we plot the Landauer transmission $T(E)$ (calculated 
from the Caroli expression eq. (\ref{eq:transm})) for $d=4a$, definitely in the tunnel regime,
and the  density of states (DOS) projected onto the tip atom of a semi-infinite
Ni chain both for  $\theta=0^\circ$ ($\rho_\|(E)$) and
$\theta=90^\circ$($\rho_\bot(E)$). The very small  transmission is dominated by
the $s$-channel, and therefore quite  independent of $\theta$.  In contrast, the
DOS is very different for $\theta=0^\circ$  and $\theta=90^\circ$. The two  peak
structure around $\epsilon_F$ for $\theta=0^\circ$  is related to the split
$E_2$  bands, which merge when $\theta=90^\circ$. In Figs. 3b and 3d we plot
the zero-bias  conductance $G(\theta)$ and the DOS at the Fermi level
$\rho(\epsilon_F,\theta)$ as a  function of $\theta$. Whereas the maximal change
in the conductance is smaller than $1\%$, the change in the DOS exceeds 200$\%$.
This challenges the simplistic link between DOS and  tunnel conductance. 

\begin{figure}
[t]
\includegraphics[width=\linewidth]{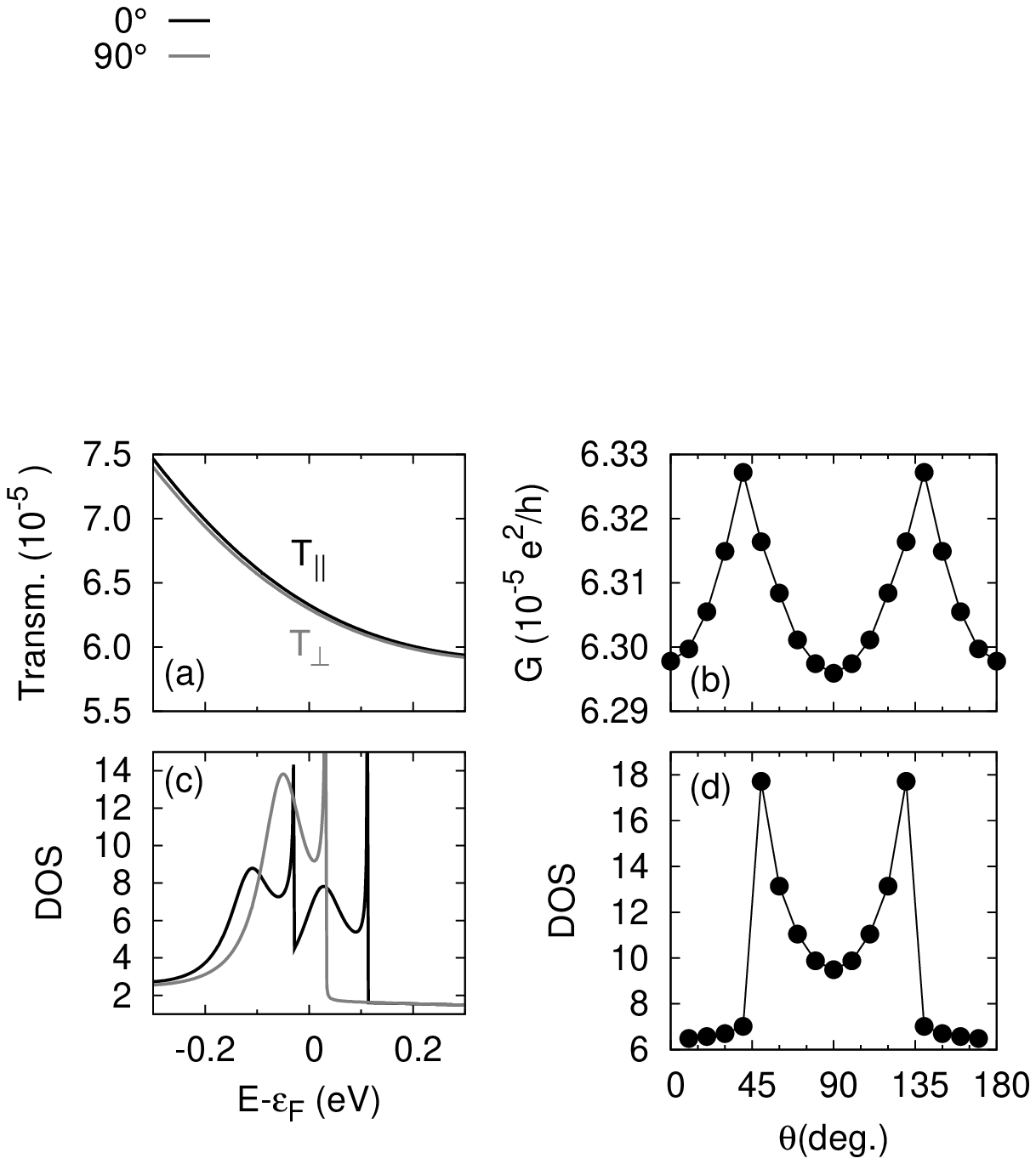}
\caption{ 
  Tunnel regime ($d=4a$):
  (a) Transmission function for magnetization angles $\theta=0^\circ$ (black) and $\theta=90^\circ$ (grey). 
  (b) Zero-bias conductance in dependence of the magnetization angle $\theta$. 
  (c) DOS projected onto tip atom as a function of energy for $\theta=0$ (black) and $\theta=90^\circ$ (grey). 
  (d) DOS projected onto tip atom at Fermi level as a function of $\theta$.
}
\end{figure}

In the tunneling regime the Landauer formula can be rewritten as (see appendix):
\begin{equation}
  \label{eq:tunnel}
  G_{\rm Tunnel}=\frac{4e^2}{h}\sum_{\alpha,\beta} |V_{\alpha\beta}|^2 \rho^L_\alpha(\epsilon_F) \rho^R_\beta(\epsilon_F)
\end{equation}
where $V_{\alpha\beta}$ is the matrix element of the Hamiltonian connecting the
$\alpha$ and $\beta$  atomic orbitals of the tip atoms of the two Ni chains and
$\rho^{L,R}_\alpha(\epsilon_F)$ is the  orbital-resolved DOS  at the Fermi
energy, i.e. the DOS projected onto an atomic orbital $\alpha$ of a tip atom.
Using this expression, the conductance calculated in Fig. 3a from the Caroli
expression is indeed nicely reproduced. 
Note, that the standard approximation by which 
the  conductance is  proportional to the product of the DOS
of the  tip atoms, 
$G\propto \sum_{\alpha}\rho^{L}_\alpha(\epsilon_F)
 \sum_{\alpha}\rho^{R}_\beta(\epsilon_F)$
is obtained only if the  $V_{\alpha\beta}$ matrix 
is  assumed to be proportional to the identity, i.e. the
tunneling matrix elements are assumed to conserve the orbital 
index and to be equal in size.  
However, this is far from being the
case when $d$ and $s$ orbitals are involved. 
In fact, in the case considered here 
the conductance  is completely dominated 
by the $V_{s,s}$ term, for which  the orbitally-resolved DOS
 $\rho^{L,R}_s$ is essentially independent of $\theta$. 
As a result, the strong dependence of the global density of states on $\theta$
is {\em not} followed, in this case, by  a strong dependence of the conductance
on $\theta$. Notice that since the transmission is dominated by the $s$
channel both the orbital polarization 
of the current and  the AMR are negligible. In general, an anisotropy in the DOS 
is not a sufficient condition to have AMR.  

The small variation of $G(\theta)$ in Fig. 3b can be traced back to the
variation of $\epsilon_F$  as a function of $\theta$ and the non-flat
$\rho_s^{L,R}(E)$. In Fig. 4a we plot $\epsilon_F(\theta)$  for a semi-infinite
Ni chain. $\Delta \epsilon_F\equiv \epsilon_F(0^\circ)-\epsilon_F(90^\circ)$
can  be as large as 10 meV. This change leads naturally to the second scenario
for enhancement of the AMR, considered in the next section:  in a situation of
resonant transport through the change of the chemical potential
$\epsilon_F$ as a function of the magnetization direction 
can result in a large  variation
of  $G$, regardless of the degree of orbital polarization of the current. It has
been recently suggested  that these resonances could arise as localized tip
states in Ni wires thicker than those considered  here \cite{Burton07}. 

\section{Sequential tunneling regime}
\label{sec:sequential}

In this section we consider a different scenario,  motivated by recent
experiments\cite{Wunderlich} and by the remarks at the end of the previous
section. We study  a single electron transistor (SET) with
Ni electrodes \cite{Ralph-Science,Liu06,Seneor07} and a non-magnetic central
island (CI)  with a discrete electronic spectrum. The CI is weakly coupled  to
the electrodes, so that the levels acquire a broadening $\Gamma$.  The position
of these levels  can be electrically tuned with a gate. Whenever a level of the
CI is in resonance with the Fermi energy the zero bias conductance of the system
has a maximum. We assume that  both  the level spacing of the CI
 states, $\Delta E$, and the charging energy $E_Q$ are much  larger than the
temperature $k_BT$ which is larger than $\Gamma$. Under  these conditions, the
system is in the Coulomb Blockade regime.

In equilibrium the chemical potential of  the central island and that of the
electrodes must be the same \cite{Been}:  $\epsilon_F(\theta)=E_C+ \epsilon_N +
eV_G$ where $N$ is the number of electrons in the  CI that satisfies  this
condition and $\epsilon_N$ is the energy  level occupied by the last electron.
>From this equation we  immediately see that the charge state of the central
island can be controlled both with the gate and with  the orientation of the
magnetization of the electrodes \cite{Wunderlich,CBAMR-theory}. This effect is 
reminiscent of the so called magneto Coulomb effect, in which the chemical
potential of the electrode is  varied with the intensity of the applied field
\cite{MCB}. Here the chemical potential is changed by rotating the applied
field.

\begin{figure}[t]
\includegraphics[width=\linewidth]{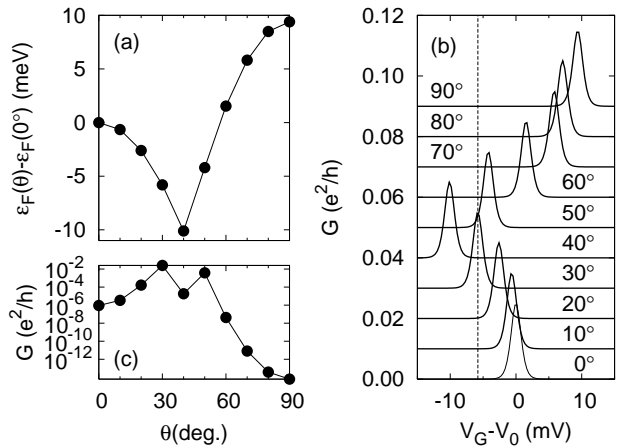}
\caption{ 
  (a) Variation of the Fermi energy of the Ni chain as a function of 
  $\theta$. (b) $G(V_G,\theta)$ for a SET device coupled to the Ni chains. 
  The curves are vertically shifted. (c) $G(5 meV,\theta)$ in logarithmic 
  scale. 
}
\end{figure}

In the $E_Q>kT>\Gamma$ situation, the linear conductance of the SET can be
obtained using either the finite temperature Landauer approach \cite{Datta}
or the sequential transport theory\cite{Been}:
\begin{equation}
  G=\frac{e^2}{h}\frac{\Gamma}{8 k_B T} \cosh^{-2}\left(\Delta/2k_B T\right),
\end{equation}
where $\Delta=E_N(V_G)+\frac{e^2}{2C}-\epsilon_F(\theta)$. In Fig. 4b we plot
$G(V_G,\theta)$ for a SET  with Ni electrodes. We take $kT=5 \Gamma=0.5$ meV.
The gate is chosen so that, for $\theta=0$ the  conductance is maximal. As
$\theta$ changes the chemical potential of the electrodes moves away from the
peak. In Fig. 4c we plot $G(\theta)$ for $V_G$ corresponding to the vertical
line in Fig. 4b. Notice the logarithmic  scale and the huge AMR, which might
have practical applications. Notice that crossing the conductance peak,  either
by gate application or magnetization rotation, implies charging the CI by one
electron  \cite{Wunderlich,CBAMR-theory}. The results of Fig. 4b assume that
$\Gamma$ is independent of $\theta$, which  is true as long as the resonant
level is not coupled to the $E_2$ and $E_1$ bands. The height of the $G(V_G)$ 
curves would depend on $\theta$ otherwise. In principle, a complete
characterization of the $G(V_G,\theta)$  curve would yield the
$\epsilon_F(\theta)$ and $\Gamma(\theta)$ functions, which would provide
valuable  information of the electronic structure of the electrodes. 

\section{Summary and conclusions}
\label{sec:summary}

We have presented  {\it ab initio} quantum transport calculations
of Ni nanocontacts as a function of the magnetization direction, $\theta$, 
going from the ballistic to the tunnel regime. We have shown that
AMR is unrelated from quantization of conductance, which is an artifact of the
scattering free calculations and  not expected in transition metal
nanocontacts.  We also show that a large variation of the density of states at
$\epsilon_F$ as a function of $\theta$  is not a sufficient condition for large AMR.
We identify two sufficient conditions to obtain largely enhanced AMR in quantum transport.
First, in the coherent regime (contact and tunneling), large AMR is related to a large degree 
of orbital polarization of the current, for a selected direction of the magnetization.
Second, in systems with resonances close to $\epsilon_F$, as it happens in single electron 
transistors with ferromagnetic electrodes, large AMR is related to a large variation of the chemical 
potential $\epsilon_F$ of the electrode as a function of $\theta$. 
We  report an {\it ab initio} calculation for this quantity.  
These findings  shed  light on the choice of materials and the design of
nanostructures with enhanced anisotropic magnetoresistance. 

We acknowledge R. Aguado, L. Brey, E. Tsymbal, E. Tosatti and C. Untiedt for
useful discussions. We acknowledge Spanish MEC and Generalitat Valenciana
for funding grants MAT2007-65487, Ramon y Cajal
Program,  GV-ACCOMP07/054 and  Consolider CSD2007-0010.

\begin{appendix}

\section{Derivation of tunneling formula}

For completeness, we derive eq. (\ref{eq:tunnel}) from the Landauer formalism in the limit 
of weak coupling between the electrodes. Eq. (\ref{eq:tunnel}) can also be obtained from 
Kubo formula (see e.g. the book by Mahan\cite{Mahan}, Sec. 9.3). Derivations similar to 
ours can be found in the literature \cite{Datta,Maekawa}.

We consider two semi-infinite electrodes L and R with atomically sharp 
tips separated by a distance $d$, as shown in Fig. \ref{fig:model}. 
We label the tip atoms of the left and right lead $0$ and $1$, respectively. 
Now the Green's function projected onto tip atom $0$ is given by:
\begin{equation}
  \label{eq:G0}
  \hat{G}_0(E) = (E-\hat{H}_0-\hat\Sigma_L(E)-\hat\Sigma_R(E))^{-1},
\end{equation}
where $\hat\Sigma_L$ is the self-energy representing the rest of the left
electrode without tip atom $0$ while $\hat\Sigma_R$ presents the self-energy of
the entire right electrode including the tip atom $1$. Thus the right
self-energy can be expressed by the Green's function of the isolated right electrode
$\hat{g}^R_1$ and  the coupling $\hat{V}$ between the left and the right tip
atom as:
\begin{equation}
  \hat\Sigma_R = \hat{V} \, \hat{g}^R_1 \, \hat{V}^\dagger.
\end{equation}
In the tunneling regime, i.e. for $d>>a$, when the coupling $\hat{V}$ becomes
very weak, the contribution  of the right self-energy to $\hat{G}_0$ can be
neglected, so that $\hat{G}_0$ becomes equal to the Green's function  of the {\it
isolated} left lead projected onto the tip atom, $\hat{g}_0^L$:
\begin{equation}
  \hat{G}_0(E) \approx (E-\hat{H}_0-\hat\Sigma_L(E))^{-1} \equiv \hat{g}^L_0(E).
\end{equation}
The Caroli expression\cite{Caroli} for the Landauer transmission through the tip atom thus
becomes:
\begin{eqnarray}
  T(E) &\approx& {\rm Tr\,}[ \hat{g}^L_0(E) \, \hat\Gamma_L(E)
   \, (\hat{g}_0^L)^\dagger(E) \, \hat\Gamma_R(E) ].
\label{T}
\end{eqnarray}
The coupling matrix of the right lead $\hat\Gamma_R$ can be re-written in terms of the
spectral function of the {\it isolated} right lead projected onto the tip atom,
$\hat{a}^R_1:=i(\hat{g}^R_1-(\hat{g}^R_1)^\dagger)$ as:
\begin{equation}
  \hat\Gamma_R:=i(\hat\Sigma_R-\hat\Sigma_R^\dagger)=\hat{V}\,\hat{a}^R_1\,\hat{V}^\dagger.
\end{equation}
The first three  terms in eq. (\ref{T}) are computed 
using the algebraic identity $\hat{g}^L_0 \, \hat\Gamma_L \,
(\hat{g}^L_0)^\dagger=i(\hat{g}^L_0-(\hat{g}^L_0)^\dagger)=\hat{a}^L_0$ 
where $\hat{a}^L_0$ is the spectral function of the {\it isolated} left lead
projected onto the tip atom $0$, we  find for the transmission in the tunneling
regime:
\begin{equation}
  T(E) \approx {\rm Tr}\,[ \hat{a}^L_0(E) \, \Gamma_R(E) ] = 
  {\rm Tr}\,[ \hat{a}^L_0(E) \, \hat{V} \, \hat{a}^R_1(E) \, \hat{V}^\dagger ].
\end{equation}
Thus the zero-bias conductance which is given by the transmission function at the Fermi energy, can be 
approximated in the tunneling regime by:
\begin{eqnarray}
  G &=& \frac{e^2}{h} \times T(\epsilon_F) \approx \frac{e^2}{h} \times {\rm Tr}\,[ \hat{a}^L_0(\epsilon_F) \, \hat{V} \, \hat{a}^R_1(\epsilon_F) \, \hat{V}^\dagger ]
  \nonumber\\
  &=& \frac{e^2}{h} \sum_{\alpha,\alpha^\prime,\beta,\beta^\prime} a^L_{\alpha\alpha^\prime}(\epsilon_F) \, V_{\alpha^\prime\beta} \, 
  a^R_{\beta\beta^\prime}(\epsilon_F) \, V^\ast_{\alpha\beta^\prime}, 
\end{eqnarray}
where in the last step we have labeled states on the left tip by $\alpha$ and $\alpha^\prime$ and states on the 
right tip by $\beta$ and $\beta^\prime$. The spectral functions $\hat{a}^L$ and $\hat{a}^R$ are diagonal in the
basis of eigenstates of the isolated left and right lead, and the diagonal elements yield the DOS projected onto 
the eigenstates: $a^L_{\alpha\alpha}=2\rho^L_\alpha$ and $a^R_{\beta\beta}=2\rho^R_\beta$ where $\alpha$ and 
$\beta$ now label the projections of the eigenstates onto the tip atoms. 
Thus we obtain eq. (\ref{eq:tunnel}):
\begin{equation}
  G \approx \frac{e^2}{h} \sum_{\alpha,\beta} \rho^L_\alpha(\epsilon_F) \, 
  V_{\alpha\beta} \, \rho^R_\beta(\epsilon_F) \, V^\ast_{\alpha\beta}.
\label{Gtun}
\end{equation}
Notice that this result relates the tunnel conductance to the product of the 
{\it orbital-resolved} DOS of the electrodes, as opposed to the {\it total} DOS.

\end{appendix}

\end{document}